\documentclass[useAMS, usenatbib]{mn2e}

\usepackage[dvips]{graphicx}
\input{epsf}

\title[Can a photometric redshift code determine extinction?]{Can a
photometric redshift code reliably determine dust extinction?}
\author[T.S.R Babbedge, R. Whitaker and S. Morris]
       {T.S.R Babbedge$^1$\thanks{Email: tsb1@imperial.ac.uk},
       R. Whitaker$^2$\thanks{Email: r.j.whitaker@durham.ac.uk},
       S. Morris$^2$\thanks{Email:
       Simon.Morris@durham.ac.uk, Visiting Astronomer,
       Canada-France-Hawaii Telescope operated by the National Research
       Council of Canada, the Centre National de la Recherche
       Scientifique de France and the University of Hawaii.}\\
        $^1$Astrophysics Group, Blackett Laboratory, Imperial College London,
        Prince Consort Road, London SW7 2BW, UK.\\
        $^2$Astronomy and Astrophysics, Physics Department, South Road, Durham DH1 3LE, UK.}
   \date{Accepted 2005 May 04.
      Received 2005 April 04;
      in original form 2004 December 21}

\pagerange{\pageref{firstpage}--\pageref{lastpage}}
\pubyear{2005}

\begin{document}
\maketitle
\label{firstpage}
\begin{abstract}

Photometric redshifts can be routinely obtained to accuracies of
better than 0.1 in $\Delta$z/(1+z).  However, the issue of dust
extinction is one that has still not been well quantified.  In
this paper the success of two template-fitting photometric
redshift codes (I{\scriptsize MP}Z and HYPERZ) at reliably returning A$_{\rm V}$
in addition to redshift is explored. New data on the CNOC2
spectroscopic sample of 0.2$<$z$<$0.7 galaxies are presented. These
data allow us to estimate A$_{\rm V}$ values from the observed
Balmer decrements. We also investigate whether the empirical value
of $\gamma=0.44$, the ratio between gas- and star- derived
extinction, as determined by \cite{Calzetti2001NewAR..45..601C},
is necessarily the best value for this sample.\\

When comparing the two codes to the Balmer-derived A$_{\rm V}$, a
correlation between the photometrically derived A$_{\rm V}$,
Phot-A$_{\rm V}$, and the Balmer-A$_{\rm V}$ is found. The
correlation is improved when the empirical value of $\gamma=0.44$ is allowed to vary.  From
least-squares-fitting the minimum in the reduced $\chi^2$
distribution is found for $\gamma\sim0.25\pm0.2$. For the sample of
galaxies here, the factor of two difference in covering factor
implied by the Calzetti ratio is found to be plausible.  The CNOC2
galaxies with detected Balmer lines have some preference for an
increased covering factor difference, which would perhaps imply
they are undergoing more rapid, `bursty' star formation than the
galaxies Calzetti used in her derivation.
\end{abstract}



\newcommand{\mnras}{MNRAS}

\newcommand{\apj}{ApJ}

\newcommand{\apjl}{ApJL}

\newcommand{\apjs}{ApJS}

\newcommand{\aj}{AJ}

\newcommand{\aap}{AAP}

\newcommand{\araa}{ARA\&A}

\newcommand{\pasp}{PASP}

\newcommand{\nat}{Nature}


\section{Introduction}

\label{sec:intro} Star formation rates (SFRs) and their global
history (SFH) form the backbone of a slew of methods
(observational, numerical, and analytical) investigating the
processes of galaxy formation and evolution over cosmic time.  The
SFH is important in indicating likely eras of dominant activity
and in providing a self--consistent picture of chemical
enrichment.  This can then be compared to the predictions of
semi--analytical models of galaxy formation and with
inter--galactic medium (IGM) absorption line diagnostics. However,
SFRs can be imprecise  due to complications arising from dust
extinction.

COBE measurements of the cosmic far--IR/sub--mm background energy
density (\citealt{Puget1996A&A...308L...5P}) showed it to be equal
to, or greater than, the UV/optical background (e.g.
\citealt{Hauser1998ApJ...508...25H}), implying that a large
fraction of the energy from stars over the history of the Universe
is hidden in the optical due to dust.  The role of dust in high
redshift galaxies has been discussed by many authors (e.g.
\citealt{RR1997MNRAS.289..490R};
\citealt{Pettini1998ApJ...508..539P};
\citealt{Calzetti1999ApJ...519...27C};
\citealt{Adelberger2000ApJ...544..218A}).  For example, star
forming galaxies detected via the Lyman break technique at
z$\sim$2 - 4 are estimated to be highly extincted in the
rest--frame UV, meaning star formation rates are $\sim3 - 10$
times higher than if dust is ignored (e.g.
\citealt*{Meurer1999ApJ...521...64M}).  Correction factors found
for other high redshift star forming galaxies are of a similar
order.  The exact form of this extinction correction remains
uncertain and in particular so does its evolution with epoch.

Hence an important improvement for optical--based SFH studies is
the determination of the extinction of galaxies, in addition to
their redshifts.  In order to fully allow for variation from
galaxy to galaxy, extinction needs to be measured as an additional
free parameter to redshift.  However the study of
\cite*{Bolzonella2000A&A...363..476B} found that the inclusion of
A$_{\rm V}$ as a free parameter in photometric redshift codes
caused significant increases in aliasing.  In a similar technique
developed in \cite{RR2003MNRAS.344...13R}, hereafter RR03, and
extended in \cite{BabbedgeI}, hereafter B04, these aliasing
problems were reduced by setting several A$_{\rm V}$ priors.\\

In this paper a sample of galaxies from the Canadian Network for
Observational Cosmology (CNOC2) Field Galaxy Redshift  Survey
(\citealt{Yee2000ApJS..129..475Y}) is used to investigate the
reliability of A$_{\rm V}$ values as determined by two SED
template fitting photometric redshift codes -- HYPERZ
(\citealt{Bolzonella2000A&A...363..476B}) and I{\scriptsize MP}Z (see B04) -- by
comparing the returned [z$_{phot}$, A$_{\rm V}$] to the
spectroscopically derived redshifts and Balmer decrement derived
A$_{\rm V}$'s as calculated from CFHT MOS spectroscopic data.\\ In
\S\ref{sec:cnoc2} the CNOC2 galaxy sample is set out, along with
the follow-up CFHT spectroscopic data and Balmer decrement
calculations.  In \S\ref{sec:balmerphotom} we discuss the link
between Balmer extinction and photometry.  The photometric
redshift method is briefly outlined in \S\ref{sec:method} and the
results of applying the two redshift codes to the CNOC2 sample are
presented in \S\ref{sec:avresults}.  Overall discussions and
conclusions are presented in \S\ref{sec:avdisc_conc}.

Note that for these investigations the flat, $\Omega_\Lambda$=0.70
cosmological model with H$_0$=72 km s$^{-1}$Mpc$^{-1}$ is used.
\section{CNOC2 overview}
\label{sec:cnoc2} 
\begin{figure}
\begin{center}
\includegraphics[width=7cm,height=8.5cm,angle=90]{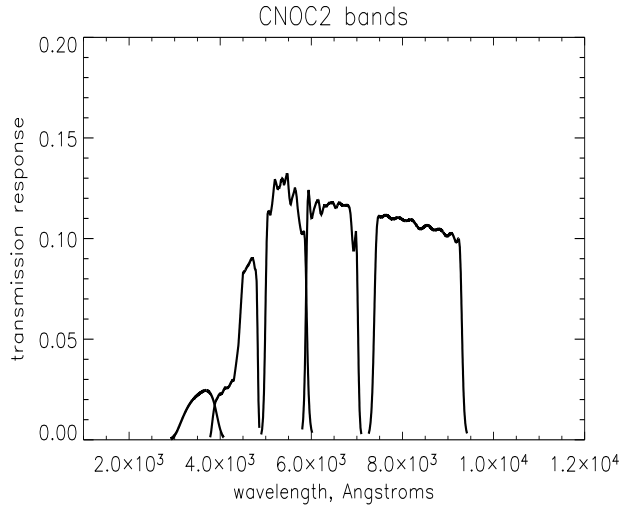}
\caption{CNOC2 passbands combined with CCD response.  From short
to long wavelength they are U, B, V, $R_c$,
$I_c$.}\label{fig:bands}
\end{center}
\end{figure}

The 2nd Canadian Network for Observational
Cosmology Field Galaxy Redshift Survey (CNOC2) was conducted over
a series of 53 nights on the Canada-France-Hawaii Telescope from
1995 -- 1998 (\citealt{Yee2000ApJS..129..475Y}). The survey covers
a total area of 1.5 sq. deg. spread over 4 patches equally spaced
in RA.  The dataset includes $\sim 6000$ galaxy spectra with
spectroscopic redshifts to a nominal limit of $R_c\sim 21.5$ in
addition to 5-colour ($I_c$, $R_c$, $V$, $B$, $U$)\footnote{$V$
photometry is actually $g$--band data
  calibrated to the $V$ system based on Landolt standards} photometry
of 40,000 galaxies, complete to $R_c$=23.0 mag. The mean limiting
magnitudes (Vega, 5$\sigma$) in each filter for the four areas are
given in Table \ref{table:limits} and the filter response curves are
shown in Figure \ref{fig:bands}.

Use of a band-limiting filter restricted the spectral window in
the survey to $4400-6300\mbox{\AA}$. This includes the [OII]
emission line for redshifts between 0.18 and 0.69. The survey's
results on galaxy clustering over this range have been published
(\citealt{Shepherd2001ApJ...560...72S};
\citealt{Carlberg2000ApJ...542...57C}) whilst an analysis of
cosmic star formation history is in preparation (Whitaker et al,
in prep.).

\begin{table}
\caption{5-sigma limits (Vega) for the four CNOC2 areas, and the
Galactic extinction $E(B-V)$ in each area.  This can be converted
to a correction in each band via Cardelli et al. (1989) with
R$_{\rm V}$=3.1.}
\begin{tabular}{|c c c c c c c|}
\hline
Area & $I_c$ & $R_c$ & $V$ & $B$ & $U$ &$E(B-V)$\\
  \hline
0223 & 22.97 & 24.02 & 23.95 & 24.55 & 22.98 & 0.036\\
1447 & 23.52 & 23.72 & 24.35 & 24.76 & 23.27 & 0.029\\
0920 & 22.85 & 24.03 & 23.94 & 24.55& 23.18 & 0.012\\
2148 & 22.69 & 23.88 & 23.70 & 24.48 & 23.08 & 0.035\\
\hline
\end{tabular}\
\label{table:limits}
\end{table}

\subsection{Extension of the CNOC2 Survey}

In this paper we present follow-up observations of a subset of the
CNOC2 sample of galaxies. The goal of the additional observations was
to observe wavelength ranges including the H$\alpha$ and H$\beta$ lines
in order to set limits on the reddening of the objects and hence to
allow estimation of the unreddened star formation rates for this
subsample. As will be shown below, this subsample then allows us to
test other means of reddening estimation and hence obtain reddening
measurements for the entire CNOC2 sample.

In order to maximise the run efficiency, masks were designed with slits
assigned as a first priority to objects with 0.2$<$z$_{spec}<$0.37
(allowing observation of both Balmer lines) and detected [OII]
$\lambda$3727 emission. Second priority was assigned to objects within
the same redshift range, but no detected [OII], and third priority to
galaxies within the CNOC2 magnitude range which did not yet have
redshifts. This last sample will not be used in this paper.

Data was taken with the CFHT MOS spectrograph \citep{Crampton1992}
during a 4 night observing run in August 1999. 17 masks containing a
total of 719 slits were observed, spread across 3 of the CNOC2
`patches'. The R300 grism was used giving a potential wavelength
coverage from 4000-10,000{\AA} (depending on slit location), with a
dispersion of ~5{\AA} per CCD pixel. A slit width of 1.5$\arcsec$ was
used giving a nominal spectral line FWHM of 3.4 pixels or 17{\AA}.

The data were reduced in IRAF
\citep{Tody1993adass...2..173T}\footnote{IRAF, the Image Reduction
and Analysis Facility, is a general purpose software system for
the reduction and analysis of astronomical data. It is written and
supported by programmers in the Data Products Program of the
National Optical Astronomy Observatory, which is operated by the
Association of Universities for Research in Astronomy, Inc., under
cooperative agreement with the National Science Foundation.}. The
reduction was based on a slightly modified version of the CNOC2
standard reductions of \cite{Yee2000ApJS..129..475Y}. The steps
included object finding, tracing and extraction, wavelength
calibration, flux calibration, and interpolation over bad sky
subtraction and zero order residuals. Error vectors were carried
through the same procedure.

\subsection{Measurement of the Balmer Decrement from CNOC2}

Our own purpose written code (Whitaker et al., in prep, based on the
code by \citealt{Balogh1999PhDT........11B}) was used to measure the
fluxes and equivalent widths of the Balmer lines in each spectrum. In
brief, the code measures the flux and equivalent width over pre-defined
spectral windows, set for each line. Each window includes two continuum
regions and a line region. We perform a $1.5\sigma$ clip on the
continuum regions to reject outliers and improve the quality of the
continuum fit. Table \ref{table:windows} shows the windows we use for
H$\alpha$ and H$\beta$ in this study. We note that our H$\alpha$
definition is similar to that of \cite{Balogh2002MNRAS.337..256B} with
two minor modifications: Firstly we use a wider line region in order to
fully encompass the H$\alpha$+[NII] flux in our measurement; this
necessarily reduces the size of the blue continuum region by 1\AA. This
should have minimal affect on the quality of the continuum fits.  The
H$\beta$ window we use is identical to that of \cite{Dressler1987AJ.....94..899D}.

While it is true that the continuum fit is obtained from only 7 (4)
independent points for H$\alpha$ (H$\beta$), the actual number of data
points used in the fit is considerably more: 23 (13) for H$\alpha$
(H$\beta$), i.e. $\sim3$ data points per resolution element. As a
result, any spurious points, from cosmic rays or the like will be
safely rejected without substantially reducing the quality of the fit. 

\begin{figure*}
\begin{center}
\includegraphics[height=8cm,width=15cm]{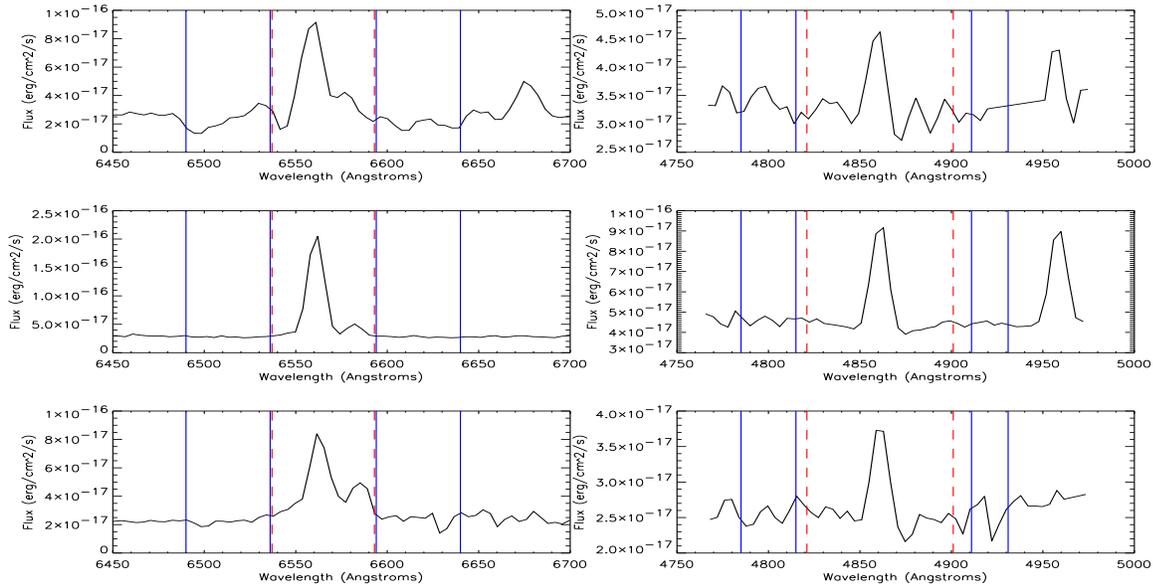}
\caption{Three typical spectra from which we measure the Balmer
  decrement. {\bf Left column} shows the H$\alpha$ region, the {\bf right column} shows the  H$\beta$.
    The window regions are overplotted: the solid (blue) lines
  show the continua regions, the dashed (red) lines the line region.
  The spectra are shown in rest-frame wavelength, the mean redshift of
  the whole sample is $\sim0.286$}\label{fig:example}
\end{center}
\end{figure*}

Figure \ref{fig:example} shows 3 typical spectra from our sample -- the
window regions are overplotted.
\begin{center}
\begin{table}
\caption{Windows used for H$\alpha$ and H$\beta$ in this study}
\begin{tabular}{c|cc}
\hline
Feature  &  \multicolumn{2}{c}{Window (\AA)}\\
 \cline{2-3}
         &  continua & line\\
\hline
H$\alpha$ &6490--6536; 6594--6640 & 6537--6593\\
H$\beta$  &4785--4815; 4911--4931 & 4821--4901\\
\hline
\end{tabular}\label{table:windows}\end{table}
\end{center}
Error estimates on the equivalent widths are computed using the
formulae set out by \cite{Bohlin1983ApJS...51..277B}. For the error on
the flux measurement we employ the following rule:

\begin{equation}
\label{eqn:fluxerror}
\sigma_{F} = F \frac{\sigma_{W}}{W}
\end{equation}

where $F$ and $W$ signify the measured flux and equivalent width and
$\sigma_{F}$ and $\sigma_{W}$ their respective errors.

The original CNOC2 dataset contains multiple observations of certain
galaxies and using these we are able to assess whether we
systematically under- or over-estimate the equivalent width errors. Our
procedure is to compute for each pair of repeats a quantity $\epsilon$
such that:

\begin{equation}
\label{eqn:epsilon}
\epsilon = \frac{W_{1}-W_{2}}{(\sigma_{1}^{2} + \sigma_{2}^{2})^{\frac{1}{2}}}
\end{equation}

Where $W$ and $\sigma$ are as in Equation \ref{eqn:fluxerror}; the
subscripts 1 and 2 refer to the separate observations. The perfect case
is where the distribution of $\epsilon$ is a unit gaussian of mean
zero. This would imply that we have accurately measured the errors and
there are no systematic under- or over-estimates.

Such a unit gaussian is not found using the raw errors and so they are
scaled according to the following equation.

\begin{equation}
\sigma_{true}^2 = A \sigma_{raw}^2 (1+B(\sigma_{raw}^2-ave^2))
\end{equation}

Where A and B are multiplying factors to be determined and $ave$ is the
average raw [OII] equivalent width error larger than 2\AA. We recompute
the distribution of $\epsilon$ for a range of A and B and check whether
it is a unit gaussian. The process is repeated until a unit gaussian is
found.

The same scaling we apply to the [OII] errors is then applied to the
H$\alpha$ errors in this work (see Whitaker et al., in prep., or
\citealt{Balogh1999PhDT........11B} for the full details of this
procedure).

\subsection{Correction For [NII] Emission and Stellar Absorption}

The window we employ in our spectral measurement code encompasses both
the H$\alpha$ and [NII] lines. We cannot separate the two lines in our
data, nor are we able to accurately de-blend them using gaussian
fitting tools (e.g. splot in IRAF) since the data is not of high enough
resolution. Instead, to correct for [NII] we assume an [NII]/H$\alpha$
ratio of 0.5 (as per \citealt{Kennicutt1992ApJ...388..310K}). This is a
simple approximation, however a by-eye examination of spectra in our
sample reveals that it is not obviously incorrect (see e.g. Figure
\ref{fig:example}). Using an extreme value for the NII/H$\alpha$ ratio
such as 0.33 which \cite{Kennicutt1983AJ.....88.1094K} find for HII
regions, results in a difference of $\sim0.4$ in A$_{\rm V}$ compared to the
[NII]/H$\alpha$=0.5 case. This value for HII regions does not
incorporate the effects of interstellar gas with its higher mean
[NII]/H$\alpha$ ratio, and since our spectra are from the galaxy as a
whole rather than individual HII regions we assume the 0.5 factor to be
the most reasonable.

The Balmer emission lines sit on top of stellar absorption due to the
presence of young and intermediate age stars in the line-emitting
galaxy, so any measurement of their fluxes must take this into account.
This is particularly true for H$\beta$, where one often finds an
emission line sitting in an absorption trough (see the middle right-hand panel
of Figure \ref{fig:example} for an example of this).

Again, due to the restrictions imposed by the data, we are not able to
reliably fit the emission and absorption components separately within
each line, so instead choose to apply the corrections used by the Sloan
Digital Sky Survey (SDSS: \citealt{Hopkins2003ApJ...599..971H}). They
measure a median absorption at H$\alpha$ of 2.6\AA (note we do not
employ their correction of 1.3\AA since we are using a window
measurement method to obtain our fluxes, not the gaussian fits of the
SDSS pipeline). For the H$\beta$ correction, we use the value for Sb
Galaxies (2\AA) given by \cite{Miller2002AJ....124.2453M}, also used
by the SDSS.

In addition to the above two corrections to the flux measurements we
also do the following: For sources with a secure H$\alpha$ detection
but either a low significance H$\beta$ detection in emission, or detection at any
significance level of H$\beta$ in absorption, we reset the
H$\beta$ flux to be its 3$\sigma$ error value. For the emission cases
this is an obvious step, for the absorption cases however, it should be
stated that the maximum H$\beta$ {\textit{emission}} flux is still the
3$\sigma$ value -- the absorption may of course be larger than this,
but since the galaxies are emitting at H$\alpha$, they must therefore
be emitting at H$\beta$ also.

\begin{figure}
\begin{center}
\includegraphics[width=8cm,height=8cm]{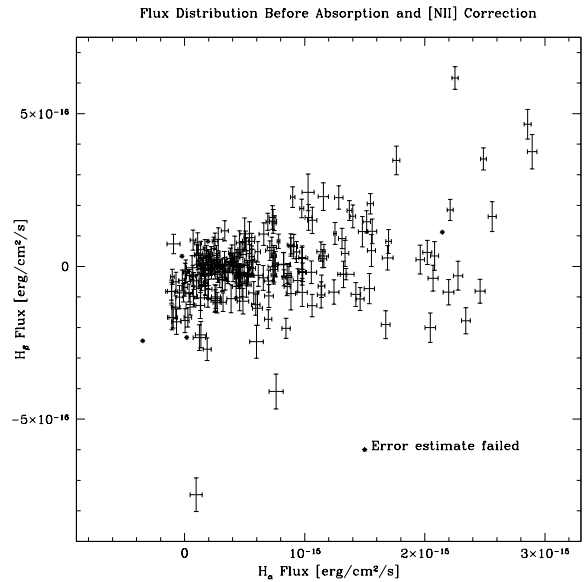}
\caption{The flux distribution of H$\alpha$
and H$\beta$ prior to corrections for [NII] and stellar
absorption.}\label{fig:before}
\end{center}
\end{figure}

\begin{figure}
\begin{center}
\includegraphics[width=8cm,height=8cm]{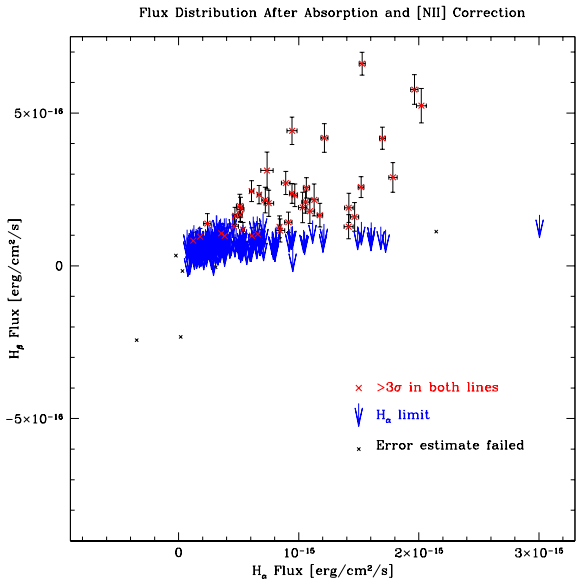}
\caption{The flux distribution of H$\alpha$
  and H$\beta$ after corrections for [NII] and stellar absorption.
  Crosses are sources with a detection of at least 3$\sigma$ in both
  H$\alpha$ and H$\beta$, arrows denote sources with an H$\alpha$
  detection of at least 3$\sigma$ but where the H$\beta$ flux has been
  reset to a 3$\sigma$ limiting value. Sources in which neither line
  could be measured at above 3$\sigma$ are not
  plotted.}\label{fig:after}
\end{center}
\end{figure}

Figures \ref{fig:before} and \ref{fig:after} show the flux distribution
of H$\alpha$ and H$\beta$ both before and after the above corrections
for [NII] and stellar absorption. In figure \ref{fig:after} the reset
H$\beta$ values are plotted as arrows, indicating an upper limit on the
flux.

\subsection{Balmer Line Detections and Limits}

We divide the 719 objects observed in the extension to the CNOC2
dataset as follows:

\begin{itemize}
\item 367 objects do not have redshifts in the original CNOC2 survey
  and so are removed from our sample.
\item Of the remaining 352 objects, 46 are removed, either due to
  skylines obscuring the H$\beta$ line (33 objects), or problems with
  the noise measurements arising from the data reduction procedures (13
  objects).
\end{itemize}

This leaves 306 objects, and of this subset:

\begin{itemize}
\item 46 objects possess both emission lines measurable at above
  $3\sigma$ accuracy (we refer to these as `measure' cases).
\item 153 objects have only H$\alpha$ measurable to an accuracy above
  $3\sigma$ -- H$\beta$ is not measurable to such accuracy in these
  spectra (we refer to these as `limit' cases, i.e. we obtain a
  lower limit to the reddening for these objects).
\item 107 objects have neither H$\alpha$ or H$\beta$ emission
  detectable at above $3\sigma$ accuracy, thus we are unable to
  estimate a lower limit on the reddening from these galaxies.
\end{itemize}

Of the set of 46 measurements, five of them fall $>3\sigma$ below
zero and of these five, four are removed (see below). This leaves
42 measured sources for the analyses carried out later in this
paper. We note that in a large enough sample, some objects are
expected to lie $3\sigma$ below zero reddening, due to the nature
of the statistics. We now describe the five objects in question:

\begin{itemize}
\item Three objects appear to be blends where two (or more) galaxies
  are co-extracted from the slit; it is impossible to derive a
  reddening measurement from these spectra.
\item One object has an error in the noise vector.
\item One object has a high quality spectrum but a relatively small
  [NII] flux. Our methodology of using a coarse correction for the
  [NII] flux serves to push extra objects into the 3$\sigma$-below-zero
  reddening range (in addition to those we naturally expect from
  gaussian statistics). This object is retained in the sample.
\end{itemize}

Of the set of 153 lower limit cases we note that 70 produce limits
below zero reddening. Again, given our error approach, this is not
surprising.

The mean redshift of the sample of `limits' and `measures' (Figure
\ref{fig:after}) is $\sim0.262$ while that of every object (Figure
\ref{fig:before}) is $\sim0.286$.

\subsection{Selection Effects}

The `measures' sample we define and use in this work is subject to
certain selection effects. 

The observations are clearly most sensitive to galaxies which have
strong emission lines and low to moderate A$_{\rm V}$. Any galaxies with weak
emission lines and moderate A$_{\rm V}$ will be lost from our sample since the
lines will have been diminished to the extent where they are
undetectable. This is also true for galaxies with large extinction --
at least H$\beta$, if not both Balmer lines, will have been diminished
by too great an amount to be detectable. The best that can be
accomplished in that case is to estimate a limit on the extinction, as
we have done here. 

The problem of skyline contamination of the H$\beta$ line is one which
all projects attempting reddening meaurements in external galaxies will
suffer from.  However this will not affect our results 
unless there exists a strong correlation of extinction with redshift such that 
certain extinctions occur at certain redshifts where the
H$\beta$ line is obscured.  Given the narrow redshift range of the survey and the 
strength of correlation that would be required this is not anticipated to be an issue.


\subsection{Calculation of extinction from the Balmer ratio}
\label{subsec:balmerav}

The H$\alpha$ and H$\beta$ Balmer lines provide a probe of
extinction for the optical region of the spectrum -- their
un-extincted intensity ratio is well known from atomic physics
\citep{Osterbrock1989} and is 2.86 in typical nebula conditions
(T=10,000K, n$_e=10^2 - 10^4$cm$^{-3}$).  Their ratio thus leads
to the Balmer optical depth, $\tau$, and can therefore provide a
measure of the dust content due to the internal reddening along
the line of sight (e.g \citealt{Calzetti1994ApJ...429..582C}),
down to an optical depth of $\tau\sim1$.

The observed ratio, R$_{obs}$, of H$\alpha$/H$\beta$ can be shown to be
related to the reddening to the gas, $E(B-V)_{g}$, according to the
following equation (e.g \citealt{Calzetti1996ApJ...458..132C}):
\begin{equation}
\label{eqn:redenning}
E(B-V)_{g}= \frac{2.5[log(R_{obs}) -
log(R_{int})]}{k(H\beta) - k(H\alpha)}
\end{equation}

where R$_{int}$ is the intrinsic H$\alpha$/H$\beta$ ratio of 2.86
and $k(H\alpha)$ and $k(H\beta)$ are the extinction coefficients
at the wavelengths of $H\alpha$ and $H\beta$ respectively (from
\citealt{Calzetti1993AAS...182.3110C}).

\begin{figure}
\begin{center}
  \includegraphics[width=8.5cm,height=13cm]{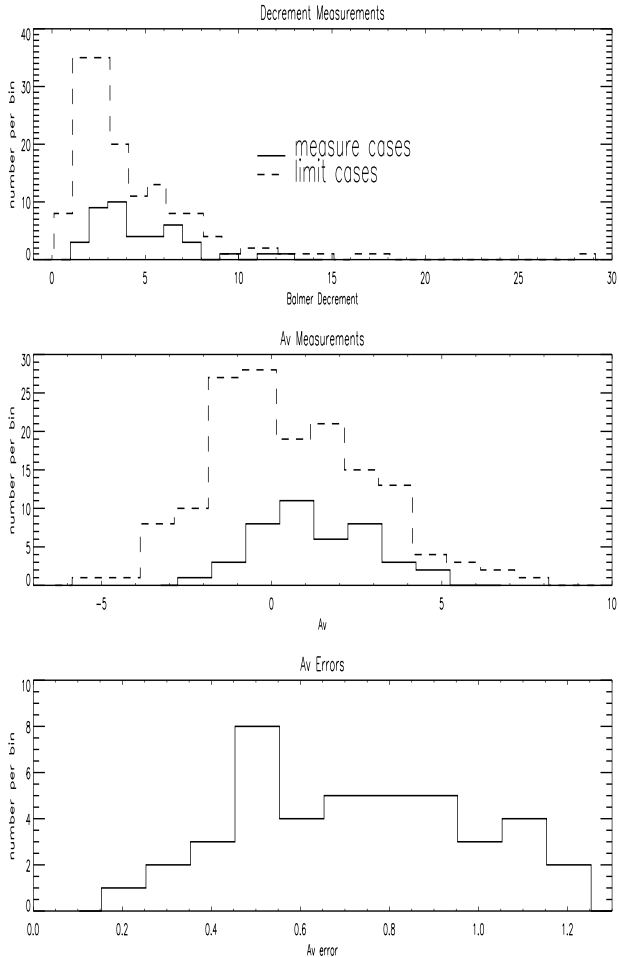} \caption{The
    distribution of Balmer decrements (top panel), Balmer-derived A$_{\rm V}$
    values (middle panel) and associated error (bottom panel). The solid line shows 
    results for the
    sources measured at above 3$\sigma$ accuracy, the dashed line the
    limit cases (in this case the decrements and Balmer-derived A$_{\rm V}$ values
    are lower-limits).
    }\label{fig:histo} \end{center} \end{figure}

The Calzetti curve (\citealt{Calzetti1993AAS...182.3110C}) was used to
compute A$_{\rm V}$'s from the Balmer decrement, with the lowest
measured A$_{\rm V}$ being -2.25 (this is the source previously mentioned
 that has low [NII] flux) and the highest 5.15.
These values are typically accurate to 30$\%$; the mean of the ratio
(A$_{\rm V}$/$\sigma$A$_{\rm V}$) is 0.27.  Figure \ref{fig:histo}
shows histograms of the 42 Balmer-measured A$_{\rm V}$ values as well
as the 153 limit cases.

The choice of extinction curve is one of the sources of
uncertainty - the extinction curves of local galaxies do differ
(e.g see review by \citealt{Calzetti2001PASP..113.1449C}).  The
largest differences are due to the strength of the feature at
$\sim2175{\rm \AA}$, thought to result from graphite grains,
amorphous carbon or PAH.  However, for the Balmer lines, this
feature is not a factor: The Balmer lines are sampling the longer
wavelength region where the extinction curves have a more constant
slope, so that differences are reduced.  The main difference
between the curves here can be parameterised by the chosen value
of the ratio R$_{\rm V}$, which is the total-to-selective
extinction ratio at V band
(\citealt{Calzetti1993AAS...182.3110C}).  In these investigations
R$_{\rm V}$=4.05 is used.

\section{Balmer extinction vs. photometry}
\label{sec:balmerphotom}
Are we able to compare photometry--derived extinction values,
hereafter Phot-A$_{\rm V}$, to Balmer-A$_{\rm V}$ values in order
to better understand the accuracy and/or reliability of
photometry--derived extinction measurements?

The crucial point here is that the Balmer decrement probes the
extinction A$_{\rm V}$ of the ionised gas in the source -- the
extinction of molecular clouds/HII regions (star-forming regions),
$E(B-V)_g$.  In the SED template-fitting method where photometry
in a number of bands is compared to a library of templates with
variable  internal A$_{\rm V}$, the resulting best-fit returns a
template type, redshift and  A$_{\rm V}$.  In this case the
extinction is due to the ISM acting on the stellar continua within
the galaxy as a whole.  Thus Phot-A$_{\rm V}$ probes the dust
obscuration of the stellar continuum, $E(B-V)_s$.  This is more
complicated than just the total amount of dust between the
observer and the source, as would be the case for a single star,
as folded into this is information on scattering and the
geometrical distribution of dust within the galaxy.

Since the Balmer and photometry methods are probing different
regimes of the galaxy extinction, they might well be expected to
give differing results. In a galaxy such as starburst where there
is a significant amount of heavily--obscured star formation the
two regimes will be linked - the galaxy's overall emission is
dominated by the young star formation.  At the opposite end of the
scale an elliptical galaxy will have little dust or ongoing SFR.
In this case the two measures are un--correlated (though both
measures are likely to be low).  For intermediate--type galaxies
the relationship falls somewhere between these two regimes.

The following relationship between the colour excess of the
stellar continuum and of the nebular emission lines is given in
\cite{Calzetti2001NewAR..45..601C} and  \cite{Calzetti97astroph}
and is as follows:
\begin{equation}
\label{eqn:calzetti}
E(B-V)_s=(0.44\pm0.03)E(B-V)_g
\end{equation}
This follows on from \cite{Calzetti1994ApJ...429..582C} where 39
starburst and blue compact galaxies were used to derive an {\it
effective} extinction law.  From this, it was found that the
difference between the optical depth to the Balmer lines, and to
the underlying stellar continuum at the lines was approximately a
factor two.  Calzetti's relationship is a purely empirical one 
but it has often, in the past, been assumed to yield the most
appropriate reddening corrections for the integrated light of
extended star-forming regions or galaxies (e.g
\citealt{Calzetti2000ApJ...533..682C};
\citealt{Westera2004A&A...423..133W}).  Here we start with this
ratio of 0.44 but then consider other possible values in order to
see how the results are affected.
\section{Photo-z Methodology}
\label{sec:method} The main outcome of applying a redshift code is
the best--fitting redshift, extinction and template SED of each
source.  In the template--fitting procedure, the observed galaxy
magnitudes are converted for each $i^{th}$ photometric band into
an apparent flux, f$_{i}^{obs}$.  The observed fluxes can then be
compared to a library of template (T) fluxes,
$f_{i}^{templ}$(z,T,A$_{\rm V}$), as calculated by convolving the
template SEDs with the filter response functions.  The reduced
$\chi^{2}$, $\chi^{2}_{red}$, is computed for each point in the
hyper-cube of redshift/template/A$_{\rm V}$ flux values and the
best--fitting solution is selected.
\subsection{The templates}
\label{subsec:templates} 
\begin{figure*}
\begin{center}
\includegraphics[width=11cm,height=15cm,angle=90]{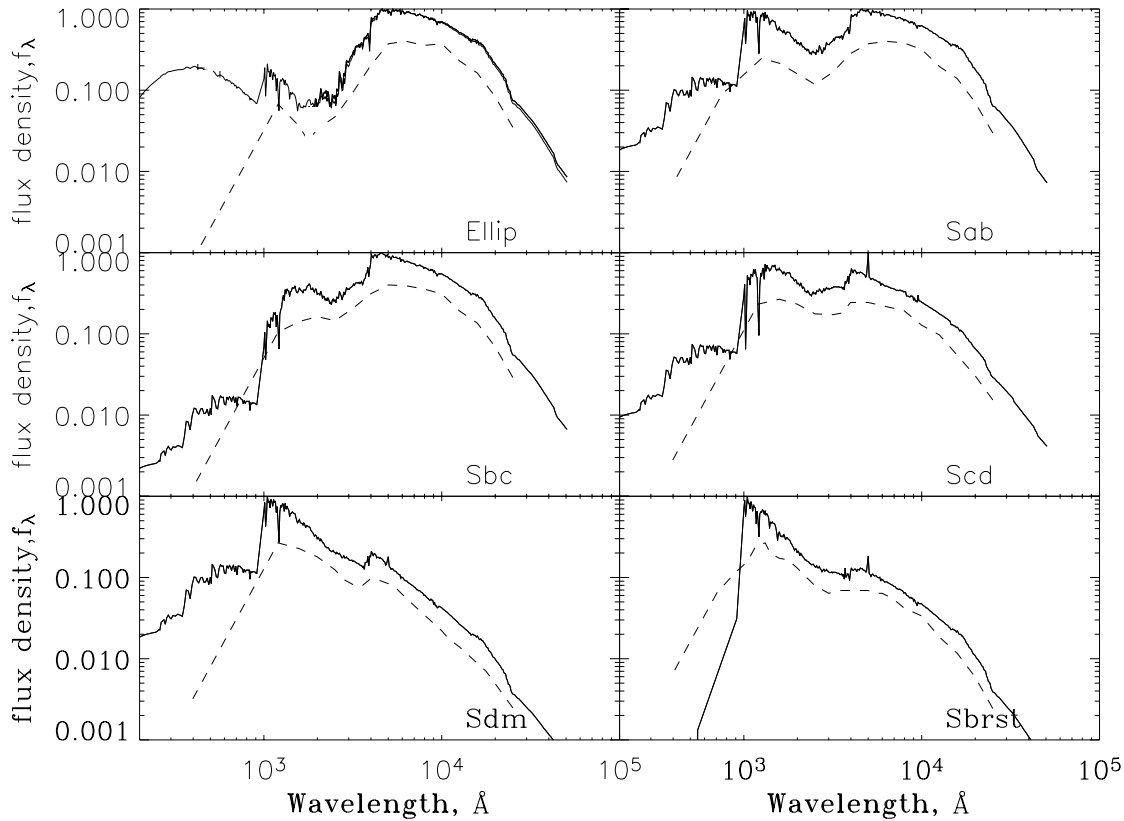}
\caption{The 6 galaxy templates used.  Dashed lines show the
original RR03 templates (offset for clarity), solid lines shows
the SSP generated versions, along with extension into the Far-UV
(sub--1000{\rm \AA}) as discussed in B04.
}\label{fig:seds}
\end{center}
\end{figure*}

Here, we use six galaxy templates,
presented in B04; E, Sab, Sbc, Scd, Sdm and starburst galaxies.
These were generated via spectrophotometric synthesis (see
\citealt{Berta2004A&A...418..913B} for more on this procedure) of
several Simple Stellar Populations (SSPs), each weighted by a
different SFR and extinguished by a different amount of dust, and
were designed to reproduce in more detail the empirical
low-resolution templates of RR03.  These SSPs have been computed
with a Salpeter Initial Mass Function (IMF) between 0.15 and 120
solar masses, adopting the \cite{Pickles1998PASP..110..863P}
spectral atlas and extending its atmospheres outside its original
range of wavelengths with \cite{Kurucz1993KurCD..13.....K} models
from 1000{\rm \AA} to 50,000{\rm \AA}, as described in
\cite*{Bressan1998A&A...332..135B}.  Nebular emission is added by
means of case B HII region models computed through the ionisation
code CLOUDY of \cite{Ferland19981998PASP..110..761F}. The adopted
metallicity is solar. In addition to the galaxy templates, two AGN
templates can be considered.  However, for the CNOC2 sample used
here it was found that the AGN templates did not provide the best
fit to any of the sources, (as expected based on their spectra) so
in this paper only the six galaxy templates are considered - they
are plotted in Figure \ref{fig:seds}.

One important consideration for the investigations here, and
template-fitting codes in general, is that the extinction output
by the photometric redshift codes refer to the Phot-A$_{\rm V}$
values of the best-fitting solutions.  However, these values are
those of the variable A$_{\rm V}$ which has been $added$ to the
template in question.  In addition to this there is also the issue
of the $inherent$ A$_{\rm V}$ of the templates.  This is
problematic in that the templates originate from empirical
templates drawn from observations, along with alterations found to
optimise redshift solutions in previous studies.  Their
re--generation via spectrophotometric synthesis where the overall
template is re-produced using a small number of simple stellar
populations (SSPs) is designed to give some insight into the
underlying physics of each template.  Hence an overall A$_{\rm V}$
can be defined, based on the A$_{\rm V}$'s of the contributing
SSPs.

In B04 these inherent A$_{\rm V}$ values were given (ranging from
0 for the elliptical to 0.74 for the starburst).  However it is
possible to reproduce the same\footnote{here, `same' means from
the point of view of flux through a set of broad-band filters}
overall template using different proportions of SSPs with
differing A$_{\rm V}$ contributions.  For example, the starburst
template can be generated from young SSPs to give the `same'
overall template, but with a total A$_{\rm V}$ of either 0.05 or
0.74.  Similarly, the Scd can be generated with an A$_{\rm V}$ of
1.6 or 0.27.  Clearly, then, we should not take the inherent
A$_{\rm V}$'s of the SSP solutions at face value (we note that in
the case of fitting SSPs to spectroscopic data most of these
degeneracies can be expected to be broken).  Empirically, we would
expect the inherent A$_{\rm V}$ of the late-type galaxies (Sab,
Sbc, Scd, Sdm) to have an extinction to the gas of perhaps
A$_{\rm V}\sim0.5-1.0$, and starbursts to be more dusty (A$_{\rm V}>1.0$,
say).

These SSP degeneracies would imply, then, that photometric data is
not able to constrain the extinction to an accuracy of less than
perhaps $\pm$1 in A$_{\rm V}$.  If this is the case, then one
would not expect a correlation to be found between the
Phot-A$_{\rm V}$ solutions and the Balmer-derived A$_{\rm V}$ even
if the two methods were probing the extinction to the same regions
of the galaxy.  The results in \S\ref{sec:avresults} will show that
this is overly pessimistic, thus implying that SSP fitting
considers more possible combinations than actually occur - there
are additional constraints imposed in reality due to the nature of
star formation, fueling and feedback that such fits do not
incorporate.
Since the details of inherent A$_{\rm V}$ are undetermined,
inclusion in the analysis would introduce an additional free
parameter.  For this relatively small sample there is insufficient
data to properly constrain this so here it is merely noted that
inherent A$_{\rm V}$ is an additional complication and that it is
expected to act to increase the Phot-A$_{\rm V}$ estimates.  If
the issue of inherent A$_{\rm V}$ is a large one then we will
expect it to swamp any Phot-A$_{\rm V}$/Balmer-A$_{\rm V}$
correlation - \S\ref{sec:avresults} will show that, in practice,
this is not the case.
\subsection{The I{\scriptsize MP}Z code}
\label{subsec:avcode} This code was presented in B04.  At high
redshift, \cite*{Massarotti2001A&A...368...74M} have shown that
correct treatment of internal (to the galaxy) dust reddening (the
Interstellar Medium) and IGM attenuation (between the galaxy and
observer) are the main factors in photometric redshift success.
The effect of internal dust reddening for each galaxy is alterable
via fitting for A$_{\rm V}$, using the reddening curve of
\cite{Savage1979ARA&A..17...73S}, and the effects of the IGM are
incorporated as in \cite{Madau1996MNRAS.283.1388M}.  Observed
fluxes are compared to template fluxes for
0.01$\leqslant$log$_{10}$(1+z)$\leqslant$0.90, equivalent to
0.02$<$z$\leqslant$6.94.

Galactic extinction in the CNOC2 regions is low; using the
extinction--wavelength relation in
\cite{Cardelli1989ApJ...345..245C} we can calculate Galactic
extinction in each of the CNOC2 bandpasses and correct for it (see
Table \ref{table:limits}), though the resulting corrections are
negligible.

For the CNOC2 sample survey the following parameters were used,
adapted from investigations presented in B04:

Templates:  The six galaxy templates (E, Sab, Sbc, Scd, Sdm and
Starburst), with IGM treatment and Galactic extinction
corrections.

Internal extinction:  A$_{\rm V}$ limits of -0.3 to 3.0 in the
A$_{\rm V}$ freedom.  Negative A$_{\rm V}$ was allowed since the
inherent A$_{\rm V}$ of the templates is non--zero.  For the
elliptical template A$_{\rm V}$ can take the value zero only, since
ellipticals are not expected to have significant internal
extinction.

Magnitude limits:  B04 found it necessary to apply
redshift--dependent absolute magnitude limits to exclude
unphysical solutions (such as super-luminous sources at high
redshift).  Here, the same limits are used:  Absolute magnitude
limits of [--22.5--2log$_{10}$(1+z)]$<$M$_B<$--13.5 for galaxies.

Prior:  I{\scriptsize MP}Z usually applies a prior expectation that the
probability of a given value of A$_{\rm V}$ declines as $|$A$_{\rm V}|$
moves away from 0.  This is introduced by minimising
$\chi^{2}_{red}$ + $\alpha$A$_{\rm V}^2$ rather than $\chi^2$
($\alpha$=2 here).  This use of a prior can be viewed as a weak implementation of
Bayesian methods and was reached at based on degeneracy studies in
B04 and RR03.  For this investigation it was of interest to turn
off this prior.  This is because the Balmer-A$_{\rm V}$ values of
the sample range from -2.25 to 5.15, or for the starlight
(multiplying by 0.44) -1 to 2.3.  Hence applying an A$_{\rm V}$
prior would naturally tend to limit the A$_{\rm V}$ solutions to
values in a similar range, regardless of the input A$_{\rm V}$
limits, and so some correlation could be achieved simply because
the sample all have Balmer-A$_{\rm V}$ values that equate to the
lower end of the A$_{\rm V}$ parameter space.

Table \ref{table:params} shows the parameters used in the final
setup for I{\scriptsize MP}Z and HYPERZ.
\subsection{HYPERZ}
\label{subsec:HYPERZ} HYPERZ is described in full in
\cite{Bolzonella2000A&A...363..476B}. Here we present an overview
of the code and the parameters we used for each run.  The HYPERZ
code also uses SED fitting in its determination of photometric
redshifts and for this work we chose to use the same templates as I{\scriptsize MP}Z.  However, for the investigations presented here, we choose
to apply HYPERZ for a `best-case' scenario where both the redshift
and template type is constrained in order to optimise the
resulting accuracy of the extinction values.  In order to create such 
a `best-case' scenario, we choose to remove one degree of freedom by
constraining the photometric redshift solution to within
($z_{spec}$/100) of the known spectroscopic redshift.
Additionally, we choose not to use the elliptical template in the
analysis presented since none of the galaxies with detected Balmer
lines are ellipticals based on visual inspection of the spectra.

We do not correct for Galactic extinction when running HYPERZ.
This correction is negligible in comparison to the photometric
errors (see Table \ref{table:limits} for extinction across the
CNOC2 patches) and would not affect the results.  Any
non-detections in a given band were replaced with a flux of zero,
the error being the limiting flux in the band (see Table
\ref{table:limits} for 5$\sigma$ limits).  Ideally, one would want to provide the 
measured flux in an aperture placed at the location of the object even when that 
measurement is fainter than the nominal flux-limit of the survey and the resulting 
value has very poor S:N.  However, the CNOC2 catalogue used here has had such 
measurements replaced by a `non-detection' flag.  Thus the best treatment in the 
template-fitting procedure is to restrict solutions to those that predict the flux in the 
non-detected band to be at or below the flux limit in that band.

The HYPERZ code was
run in its standard form, with the flat, $\Omega_\Lambda$=0.70
cosmological model with H$_0$=72km s$^{-1}$Mpc$^{-1}$ used
elsewhere in this paper and variable A$_{\rm V}$ in the range
$0<$A$_{\rm V}<3$.  Table \ref{table:params} shows the parameters
used in the final setup for I{\scriptsize MP}Z and HYPERZ.

\begin{table*}
\caption{\scriptsize{Final parameters for the two photometric redshift codes.}\label{table:params}}
\begin{center}
\begin{tabular}{|c c c |}
\hline
 & I{\scriptsize MP}Z & HYPERZ\\
 \cline{2-3}
 Templates & E, Sab, Sbc, Scd, Sdm, Sb & Sab, Sbc, Scd, Sdm, Sb \\
 M$_B$ limits & [--22.5--$2log_{10}(1+z)]<$M$_B$$<$--13.5 & not constrained (but set by z$_{spec}$) \\
 A$_v$ limts, A$_v$ step & -0.3$\le$A$_v\le$3.0, step 0.1 & 0.0$\le$A$_v\le$3.0, step 0.27 \\
 Reddening law & Cardelli 1989 & Calzetti 2000\\
 Galactic extinction correction? & Yes & No \\
 Constrained to z$_{spec}$? & One run unconstrained, one run constrained & Yes \\
 Cosmology & $\Omega_0$=1, $\Omega_\Lambda$=0.70, H$_0$=72km s$^{-1}$Mpc$^{-1}$ & $\Omega_0$=1, $\Omega_\Lambda$=0.70, H$_0$=72km s$^{-1}$Mpc$^{-1}$ \\
\hline
\end{tabular}\
\end{center}
\end{table*}

\section{Results}
\label{sec:avresults}
\subsection{I{\scriptsize MP}Z redshifts}
\label{subsec:avresultsz}
\begin{figure*}
\begin{center}
\includegraphics[width=17cm,height=7cm]{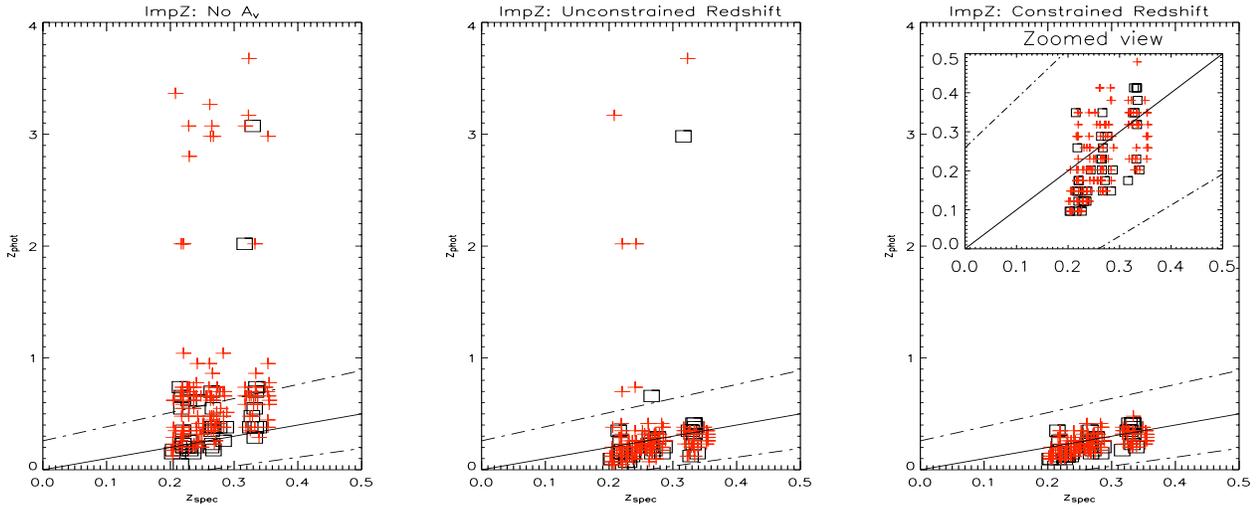}
\caption{ {\bf Photometric redshifts from I{\scriptsize MP}Z}.
Comparison of spectroscopic and I{\scriptsize MP}Z-derived redshifts for `measure' cases 
(black squares) and `limit' cases (red crosses): {\bf Left panel} shows results when A$_{\rm V}$ freedom is not considered; {\bf middle panel} shows results for unconstrained
redshift and A$_{\rm V}$ space; {\bf right panel} shows results for redshift space constrained to be within 
 0.05 in log(1+z$_{phot}$) of z$_{spec}$, but A$_{\rm V}$ unconstrained.
Dot-dashed lines denote an accuracy of 0.1 in log(1+z), a typical photometric redshift accuracy.
}\label{fig:photz}
\end{center}
\end{figure*}

First, it is of interest to see how successful the photometric redshifts are when there is no 
A$_{\rm V}$ freedom - that is, A$_{\rm V}$=0 in the solutions.  The results of this are 
plotted in the left panel of Figure \ref{fig:photz}, for `measure' cases 
(black squares) and `limit' cases (red crosses).  It is immediately clear that not allowing A$_{\rm V}$ freedom 
has caused many of the I{\scriptsize MP}Z solutions to be incorrect - the code has been forced to 
substitute (incorrectly) additional redshift in place of the reddening action of dust.  Indeed, there are a 
total of 11 of the 42 `measure' and 70 of the 153 `limit' sources (48\% of the sample) whose photometric 
redshifts lie outside the log(1+z$_{spec})\pm$ 0.1 boundaries.

If we now run I{\scriptsize MP}Z in the optimal manner, as discussed in \S\ref{subsec:avcode}, so that now 
A$_{\rm V}$ freedom is allowed, the solutions dramatically improve.   I{\scriptsize MP}Z was highly successful at deriving photometric redshifts in close agreement with the 
spectroscopic values for 40 of the 42
`measure' and 147 of the 153 `limit' sources (95\% of the sample).  Note we consider redshift
solutions in the range [0:7]).  There were 8 (2 `measure' and 6 `limit') 
sources that obtained solutions at incorrect redshifts due to degeneracies 
in [z, template, A$_{\rm V}$] space.  We measure the accuracy of the
photometric redshifts via the $rms$ scatter $\sigma_z$, calculated
as follows:
\begin{equation}
\label{eqn:rms}
\sigma_{\rm z}^2=  \sum\left(\frac{z_{phot}-z_{spec}}{1+z_{spec}}\right)^2/N
\end{equation}
with N being the number of sources with both spectroscopic
redshifts and photometric redshifts.  For this data, the $rms$,
$\sigma_z$=0.32 when including the 8 outliers, or 0.07 when they are excluded.  The 
comparison of spectroscopic and I{\scriptsize MP}Z redshifts 
is plotted in the middle panel of Figure \ref{fig:photz}.

I{\scriptsize MP}Z is therefore successful at returning accurate redshifts.  
However, can it also provide a measure of the extinction
compatible with that implied by the Balmer decrement?  Since 2 of the 42 `measure' sources 
obtain an incorrect photometric redshift this also means their extinction values are likely to be 
incorrect.  In order to remove this (small) source of error in the following comparisons to the 
Balmer-derived extinction, we now constrain the redshift range explored by I{\scriptsize MP}Z to 
lie within 0.05 in log(1+z$_{phot}$) of z$_{spec}$ (plotted in the right panel of Figure
 \ref{fig:photz}).  Now, good solutions are found for 
all 42 `measure' cases and 153 `limit' cases, with $\sigma_z$=0.06.  
It is the resulting I{\scriptsize MP}Z A$_{\rm V}$ values from this setup that are considered from 
now on in the investigation.
\subsection{Range of I{\scriptsize MP}Z A$_{\rm V}$ allowed values}
\label{subsec:avspiresults}
Although we wish to explore the accuracy of the extinction output from redshift codes by 
comparing to a sample with Balmer-derived measurements, we can also obtain an internal 
estimate of how well constrained the [z,template,A$_{\rm V}$] solution is from the reduced 
$\chi^2$ distribution.  For the solution with the minimum $\chi^2$, $\chi^2_{min}$, the question 
can be asked: `What range of A$_{\rm V}$ produces a fit at or near the correct redshift 
(within 0.05 of log[1+z$_{spec}$]), 
with a reduced $\chi^2$ within $\chi^2_{min}$+1?'

\begin{figure*}
\begin{center}
\includegraphics[width=17cm,height=6.5cm]{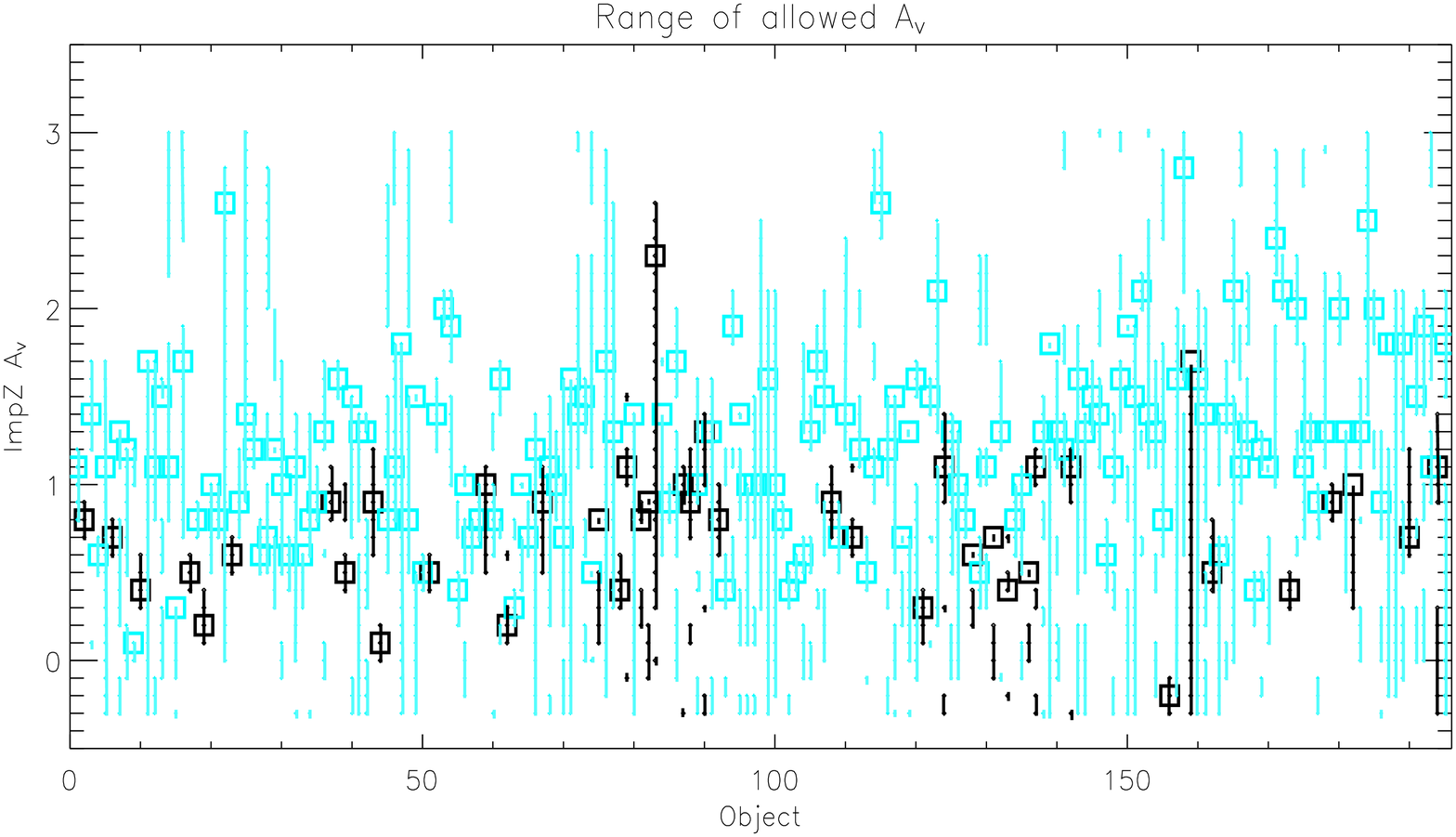}
\caption{ {\bf I{\scriptsize MP}Z A$_{\rm V}$ allowed values}.
The range of A$_{\rm V}$ parameter space for each source that provides a solution with a 
reduced $\chi^2$ within $\chi^2_{min}$+1 and that is at or near the correct redshift
 (within 0.05 of log[1+z$_{spec}$]).  `Measure' sources are shown as black lines, 
 `limit' sources as cyan lines.  The best A$_{\rm V}$ solution value is indicated 
 as a black (`measure' case) or cyan (`limit' case) square.
}\label{fig:av_range}
\end{center}
\end{figure*}

  The results of asking this question of each 
source is illustrated in Figure \ref{fig:av_range}.  It can be seen that for the majority of `measure' 
cases (black), the A$_{\rm V}$ is quite well constrained, in most cases to within 0.3 or so in A$_{\rm V}$.
  Note that the lines in this plot can be discontinuous - for example, object 150 in the plot has a 
  best-fitting A$_{\rm V}$ of 1.9 but has reasonable solutions in the A$_{\rm V}$ range -0.3 to 0.4
   and 0.8 to 1.1 which arise from fitting two other templates to the source.  It is of interest to note
    that whereas 57\% of `measure' cases do not have discontinuous solutions, this is only true 
    for 43\% of the `limit' (cyan) cases.  Thus, determining the A$_{\rm V}$ via photometry for these 
    sources is more problematic, just as it is via the Balmer ratio method.  Considering the `measure'
     cases in more detail, there are two objects where the A$_{\rm V}$ is poorly constrained (objects 
     83 and 159 in the Figure, evidenced by their long black lines).  Their best-fitting A$_{\rm V}$ values 
     are, respectively, 2.3 and 1.7, making them the most heavily extincted of the I{\scriptsize MP}Z `measure' cases.
  Comparison to their Balmer-derived A$_{\rm V}$ ($3\pm 1$ and $-1.6\pm 0.5$ respectively) would support the
   result for the first source but the Balmer decrement for the second source would imply negative, or
zero extinction, disfavouring the I{\scriptsize MP}Z best-fitting result or resulting in the interpretation
that this source is problematic.

\begin{figure}
\begin{center}
\includegraphics[width=8.5cm]{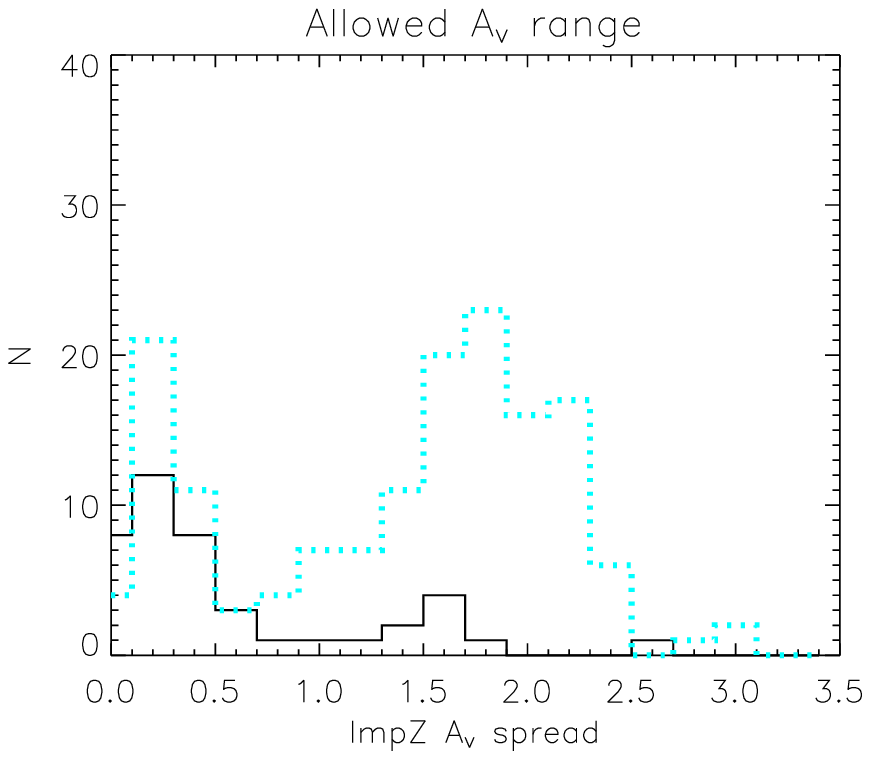}
\caption{ {\bf Histogram of the width of I{\scriptsize MP}Z A$_{\rm V}$ allowed values}.
  Distribution of the width of A$_{\rm V}$ parameter space 
  (defined by the minimum and maximum allowed A$_{\rm V}$) 
that lies within +1 of $\chi^2_{min}$.  `Measure' sources are shown as a
 black line, `limit' sources as a cyan dotted line.
}\label{fig:hist_avspread}
\end{center}
\end{figure}

 Figure \ref{fig:hist_avspread} illustrates the width of A$_{\rm V}$ parameter space 
that lies within +1 of $\chi^2_{min}$ by plotting the distribution of this `width' value (here the `width' 
is simply defined as the maximum allowed A$_{\rm V}$ minus the minimum allowed A$_{\rm V}$).  
For `measure' cases (black) the distribution drops quite steeply with width such that more than 
half the sources have a width of 0.4 or less.  There is then a slight tail made up primarily of
sources which had discontinuous solutions (such as one template with low A$_{\rm V}$ and another
 template with higher A$_{\rm V}$) whilst the two sources with poorly constrained A$_{\rm V}$ 
can be seen as a spike towards the maximum 
width of 3.3 (i.e the full -0.3 to 3 A$_{\rm V}$ range).  The distribution for the `limit' cases is more clearly 
bimodal, with a similar set of reasonably well-constrained sources with an A$_{\rm V}$-width of 0.4 or less, 
but a much larger set of sources with poorly constrained A$_{\rm V}$.  Again, this is likely to be due to 
the nature of these `limit' sources for whom the dust extinction is hard to determine (either via the Balmer lines 
or photometry).

This analysis suggests that the photometric redshift solution has an inherently low 
extinction precision (at least with five band photometry), such that A$_{\rm V}$ is only 
precise to perhaps 0.3 for most sources, and is poorly constrained for a small subset.
  Rather than defining this internally estimated width value as the error in the Phot-A$_{\rm V}$ 
  value we choose instead to take the opposite approach for the comparison with the 
  Balmer-derived A$_{\rm V}$.  We will take the Phot-A$_{\rm V}$ at face 
  value and use the supposed relation 
  to the Balmer-derived A$_{\rm V}$ to provide an external estimate of the precision of the 
  Phot-A$_{\rm V}$ measurements.
\subsection{I{\scriptsize MP}Z - comparison to Balmer}
\label{subsec:aviresults}
\begin{figure*}
\begin{center}
\includegraphics[width=15cm,height=7cm]{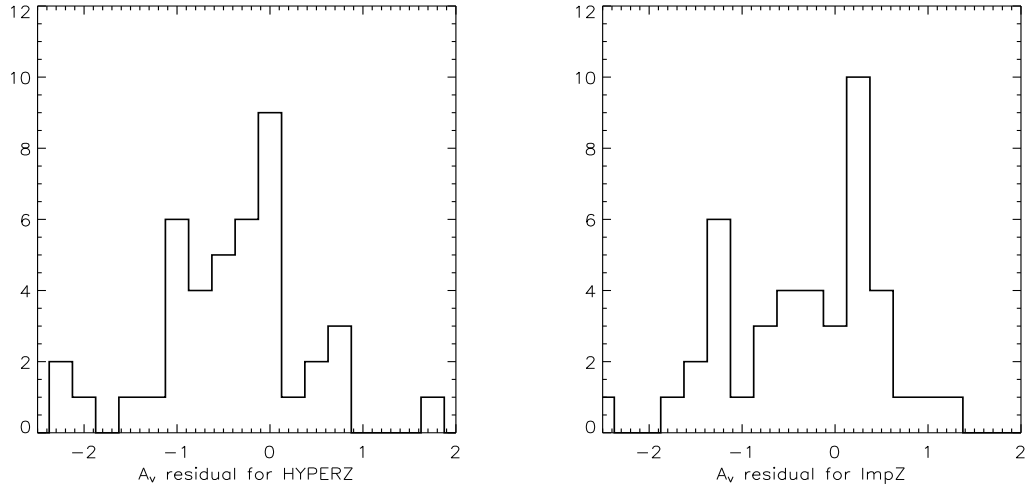}
\caption{ {\bf A$_{\rm V}$ residuals for HYPERZ (left panel) and I{\scriptsize MP}Z (right panel)}.
The residual is (0.44*Balmer[A$_{\rm V}$]-Phot[A$_{\rm V}$]).
}\label{fig:avresid}
\end{center}
\end{figure*}

We can test how well the Calzetti ratio holds by comparing the Phot-A$_{\rm V}$
 to 0.44*Balmer-A$_{\rm V}$.  The distribution of A$_{\rm V}$ 
 residuals for I{\scriptsize MP}Z is plotted in
Figure \ref{fig:avresid}.  It can be seen that there is quite a
spread to the distribution, though it is broadly centred on zero.
  Based on the findings in \S\ref{subsec:avspiresults} on the 
  precision of the Phot-A$_{\rm V}$ solutions some of this spread can be 
  expected to arise from this low precision.  Some of it can also
  be attributed to the accuracy of the Balmer-derived A$_{\rm V}$,
   which is typically accurate to around 30$\%$.

\begin{figure*}
\begin{center}
\includegraphics[width=15cm,height=7cm]{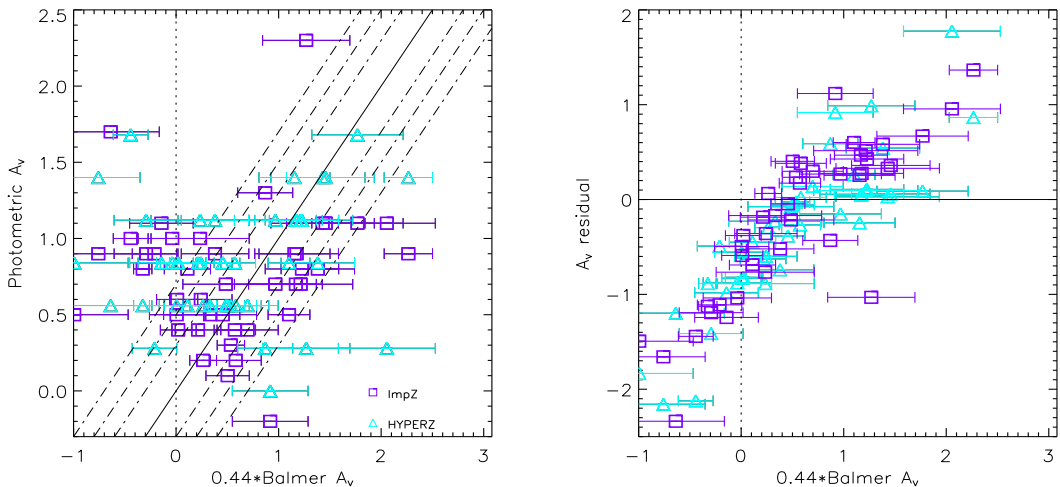}
\caption{ {\bf Left:} A$_{\rm V}$ results for HYPERZ (cyan
triangles) and I{\scriptsize MP}Z (purple squares).  Plot is the Balmer[A$_{\rm
V}$] (multiplied by the Calzetti (2001) factor of 0.44) versus the
photometrically--derived A$_{\rm V}$ .  Dot--dashed lines denote
residuals of 0.3, 0.5 and 0.7 in A$_{\rm V}$.  Errors are not
defined for the Phot-A$_{\rm V}$ values.  {\bf Right:} A$_{\rm V}$
residuals for HYPERZ (cyan triangles) and I{\scriptsize MP}Z (purple squares).
Plot is Balmer-A$_{\rm V}$ (multiplied by the Calzetti 2001 factor
of 0.44) versus the residual (0.44*Balmer[A$_{\rm
V}$]-Phot[A$_{\rm V}$]).
}\label{fig:av_comp}
\end{center}
\end{figure*}

Figure \ref{fig:av_comp} plots (purple squares) the I{\scriptsize MP}Z A$_{\rm
V}$ values and residuals as a function of Balmer-A$_{\rm V}$
(multiplied by the \citealt{Calzetti2001NewAR..45..601C} factor of
0.44).  Note that the I{\scriptsize MP}Z  A$_{\rm V}$ values have been taken at face-value and
 have not had an error assigned to them, since we wish to derive an error based on 
 the comparison to the Balmer-A$_{\rm V}$'s.

  It can be seen that the residuals are smallest for Phot-A$_{\rm V}$ values
   of around 0.5 to 1 and that the residuals increase as we move away from 
   this region (to either higher A$_{\rm V}$ or lower/negative A$_{\rm V}$).  Hence, 
   the correlation to Balmer-derived A$_{\rm V}$ appears to be best for sources 
   of intermediate extinction.  It is also clear that none of the sources that were calculated 
   as having negative A$_{\rm V}$ based on their Balmer lines obtain similar Phot-A$_{\rm V}$ 
   values, lending weight to the supposition that the Balmer method has fallen 
   down for these sources due to limitations in the technique.  
   The source with the largest Balmer-A$_{\rm V}$,
of 5.15$\pm$0.5, is also the source with the largest residual (of sources with non-negative A$_{\rm V}$).  Being so extincted, it is likely that this object is quite extreme,
so disagreement between the star- and gas-derived extinction
measures is to be expected.

Calculating the following statistics:

\begin{equation}
\label{eqn:del_avmean}
\overline{\Delta A_{\rm V}}= \sum\left(Balmer[A_{\rm V}]-Phot[A_{\rm V}]\right)/N
\end{equation}
and
\begin{equation}
\label{eqn:avrms}
\sigma_{\rm A_{\rm V}}^2=   \sum\left(Balmer[A_{\rm V}]-Phot[A_{\rm V}]\right)^2/N
\end{equation}
and outliers are defined by
\begin{equation}
\label{eqn:avout}
|\Delta A_{\rm V}|>0.5
\end{equation}

gives $\overline{\Delta A_{\rm V}}=-0.2$, $\sigma_{\rm A_{\rm V}}$=0.83 and an
outlier fraction, $\eta$, of 50 per cent.  These results are
better than one would infer from the large extinction degeneracies
seen in the SSP fitting which were discussed in
\S\ref{subsec:templates}.  Thus, a relationship between
Balmer-derived and photometry-derived extinction measures is
obtainable, though not a strong one.
\subsection{HYPERZ - comparison to Balmer}
\label{subsec:avhresults} Applying HYPERZ in a `best-case'
configuration (constraining the HYPERZ redshift solutions to the
spectroscopic values and excluding the elliptical template) gives similar statistics of 
$\overline{\Delta A_{\rm V}}=-0.3$, $\sigma_{\rm A_{\rm V}}$=0.84 and  an
outlier fraction, $\eta$, of 50 per cent.  The distribution of
A$_{\rm V}$ residuals for HYPERZ is plotted in Figure
\ref{fig:avresid}.  As with I{\scriptsize MP}Z, the distribution is broad but reasonably 
centred on zero.

Figure \ref{fig:av_comp} plots (cyan triangles) the HYPERZ A$_{\rm
V}$ values and residuals as a function of Balmer-A$_{\rm V}$
(multiplied by the \citealt{Calzetti2001NewAR..45..601C} factor of
0.44).  The residuals are again
correlated with the Balmer-A$_{\rm V}$, being smallest for Phot-A$_{\rm V}$ values
 of around 0.5 to 1, and the negative Balmer-A$_{\rm V}$ sources are again 
 in poor agreement.
   
Thus, HYPERZ and I{\scriptsize MP}Z portray a similar correlation to the Balmer-A$_{\rm V}$, 
though the agreement is noisy.
\subsection{Comparison of I{\scriptsize MP}Z and HYPERZ}
\label{subsec:phot_comp}
\begin{figure}
\begin{center}
\includegraphics[width=8cm]{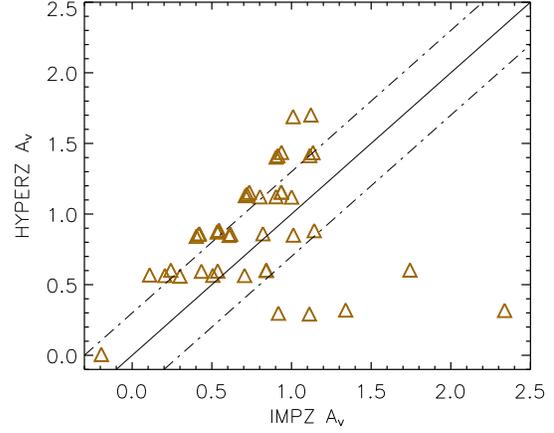}
\caption{ {\bf Comparison between I{\scriptsize MP}Z and HYPERZ:} A$_{\rm V}$
results for I{\scriptsize MP}Z versus those from HYPERZ.  Solid line is exact
agreement, dot-dashed lines are residuals of 0.3 in A$_{\rm V}$.  Note that for plotting purposes the values have been randomly altered by up to 0.02 in the x and y directions in order to separate points with the same/very similar values.
}\label{fig:imp_hyp}
\end{center}
\end{figure}

As well as comparing the two photometric redshift code's
extinction outputs to the Balmer-derived values, it is instructive
to compare them to one another to see if they tend to agree on a
similar extinction value for a given source.  A plot of I{\scriptsize MP}Z-A$_{\rm V}$ versus HYPERZ-A$_{\rm V}$ is given on Figure
\ref{fig:imp_hyp}.

This shows that the two codes are in reasonable agreement about
the extinction of a given source.  Twenty-five of the forty-two
sources (60\%) agree within $<0.4$ in A$_{\rm V}$,  thirty-five
sources (83\%) agree within $<0.5$.  The main difference appears
to be for five sources where I{\scriptsize MP}Z gives a high value of A$_{\rm V}>0.8$
whilst HYPERZ tends to return a smaller A$_{\rm V}$ estimate.  Two
of these sources (this includes object 159 mentioned in \S\ref{subsec:avspiresults}) 
have Balmer decrements that imply negative, or
zero extinction, favouring the HYPERZ result or the interpretation
that the sources are problematic.  Two others have intermediate
Balmer-A$_{\rm V}$ consistent with either code's results and one has 
a larger Balmer-A$_{\rm V}$ (this is object 83 mentioned in \S\ref{subsec:avspiresults}) 
thus favouring the I{\scriptsize MP}Z result.

Calculating similar statistics as when comparing to the
Balmer-A$_{\rm V}$, comparison between the two codes' A$_{\rm V}$
values gives $\overline{\Delta A_{\rm V}}=-0.09$, $\sigma_{\rm A_{\rm V}}$=0.53
and  an outlier fraction, $\eta$, of 17 per cent.  This internal
consistency check between the two codes gives increased confidence
in the photometric redshift template-fitting method as a technique
to obtain extinction.
\subsection{The Calzetti ratio}
\label{subsec:cal_ratio}
\begin{figure}
\begin{center}
\includegraphics[width=8cm]{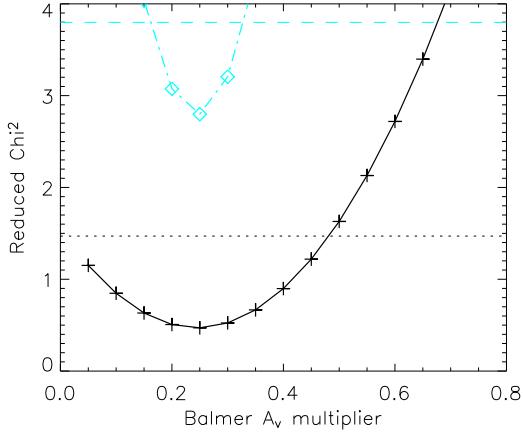}
\caption{ {\bf $\chi^2$ analysis}.  The reduced $\chi^2$ for I{\scriptsize MP}Z
(solid line with crosses) and HYPERZ (dot-dashed line, diamonds)
as a function of $\gamma$, the chosen ratio between
photometrically-derived and Balmer ratio-derived extinction
measures.  The $\chi^2$ values that are 1 above the minimum in the
two distributions are indicated by horizontal lines (I{\scriptsize MP}Z, dotted;
HYPERZ, dashed).}\label{fig:balmer_chi}

\end{center}
\end{figure}

We now consider the empirical ratio of 0.44 introduced by
\cite{Calzetti97astroph} and set out in Eqn. \ref{eqn:calzetti}.
We consider a range of other ratios, $\gamma$, so that
\begin{equation}
\label{eqn:calzetti_x}
E(B-V)_s=\gamma E(B-V)_g
\end{equation}
Varying this ratio from 0.05$<\gamma<1$, a $\chi^2$ minimisation
analysis is carried out in order to find the ratio that gives the
best correlation between the Phot-A$_{\rm V}$ and the
Balmer-A$_{\rm V}$, weighted by the errors in the Balmer-A$_{\rm
V}$ measurements.  The results of this analysis are shown in
Figure \ref{fig:balmer_chi}.  For both I{\scriptsize MP}Z and HYPERZ, the lowest
reduced $\chi^2$ is reached for $\gamma\sim0.25$.  Based on the
change in $\gamma$ that increases the reduced $\chi^2$ by one, the
1$\sigma$ error on this value is $\gamma=0.25\pm0.25$ for I{\scriptsize MP}Z and
$\gamma=0.25\pm0.1$ for HYPERZ.\\

\begin{figure*}
\begin{center}
\includegraphics[width=15cm,height=7cm]{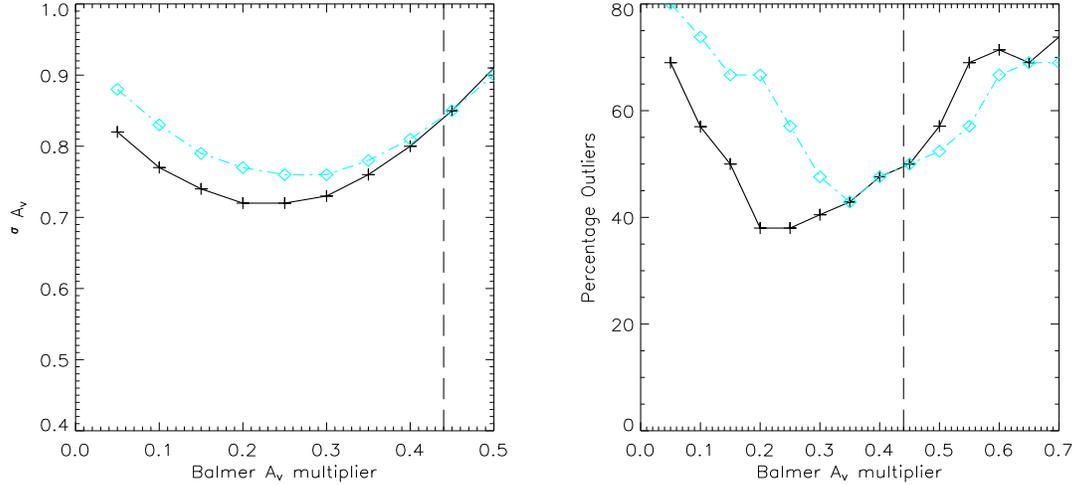}
\caption{ {\bf Left panel:} $\sigma_{\rm A_{\rm V}}$ for I{\scriptsize MP}Z (solid line with
crosses) and HYPERZ (dot-dashed blue line, diamonds) as a function
of $\gamma$, the chosen ratio between photometrically-derived and
Balmer ratio-derived extinction measures.  The Calzetti value of
$\gamma=0.44$ is indicated as a long-dashed line.
{\bf Right panel:} Percentage outliers for I{\scriptsize MP}Z (solid line with crosses)
and HYPERZ (dot-dashed blue line, diamonds) as a function of
$\gamma$, the chosen ratio between photometrically-derived and
Balmer ratio-derived extinction measures.  The Calzetti value of
$\gamma=0.44$ is indicated as a long-dashed line.
}\label{fig:balmer_gamma_stat}
\end{center}
\end{figure*}

The resulting statistical measures, $\sigma_{\rm A_{\rm V}}$ and the
outlier fraction, $\eta$, are also plotted as a function of
$\gamma$ for the I{\scriptsize MP}Z and HYPERZ results (Figure
\ref{fig:balmer_gamma_stat}).  
The left panel shows how the $rms$ in
the residual, $\sigma_{\rm A_{\rm V}}$, varies with $\gamma$.  A clear
minimum is seen at $\gamma\sim$ 0.15 to 0.35 for I{\scriptsize MP}Z results, and
at around 0.2 to 0.4 for HYPERZ.
A similar minimum is seen in the range $\gamma\sim$ 0.2 to 0.35
for I{\scriptsize MP}Z results when the outlier fraction, $\eta$, is plotted
against $\gamma$ in the right-hand panel.  For
HYPERZ, the minimum is at around $\gamma\sim$ 0.3 to 0.45.\\

This analysis suggests that for this sample, the Calzetti ratio of
0.44 is a reasonable choice for $\gamma$ within the accuracy of
the method, though a value of $\sim0.25$ is preferred.  A choice of  $\gamma$=0.25 gives
the following statistics:

For I{\scriptsize MP}Z; $\overline{\Delta A_{\rm V}}=-0.4$, $\sigma_{\rm A_{\rm V}}$=0.72
and an outlier fraction, $\eta$, of 38 per cent.

For HYPERZ; $\overline{\Delta A_{\rm V}}=-0.5$, $\sigma_{\rm A_{\rm V}}$=0.76, $\eta$, of 57 per cent.
\begin{table}
\caption{Statistical results for the different comparisons between
Balmer-A$_{\rm V}$ and Phot-A$_{\rm V}$.}
\begin{tabular}{|c | c c c |}
\hline
& $\overline{\Delta A_{\rm V}}$ &  $\sigma_{\rm A_{\rm V}}$ & outlier fraction, $\eta$\\
 \cline{2-4}
I{\scriptsize MP}Z, $\gamma$=0.44 &-0.2 & 0.83& 50\% \\
HYPERZ, $\gamma$=0.44 & -0.3& 0.84& 50\% \\
 \cline{2-4}
I{\scriptsize MP}Z, $\gamma$=0.25 &-0.4& 0.72&38\%  \\
HYPERZ, $\gamma$=0.25 &-0.5 & 0.76& 57\% \\
\hline
I{\scriptsize MP}Z/HYPERZ comparison &-0.09 & 0.53& 17\% \\
\hline
\end{tabular}\
\label{table:stats}
\end{table}

\begin{figure*}
\begin{center}
\includegraphics[width=15cm,height=7cm]{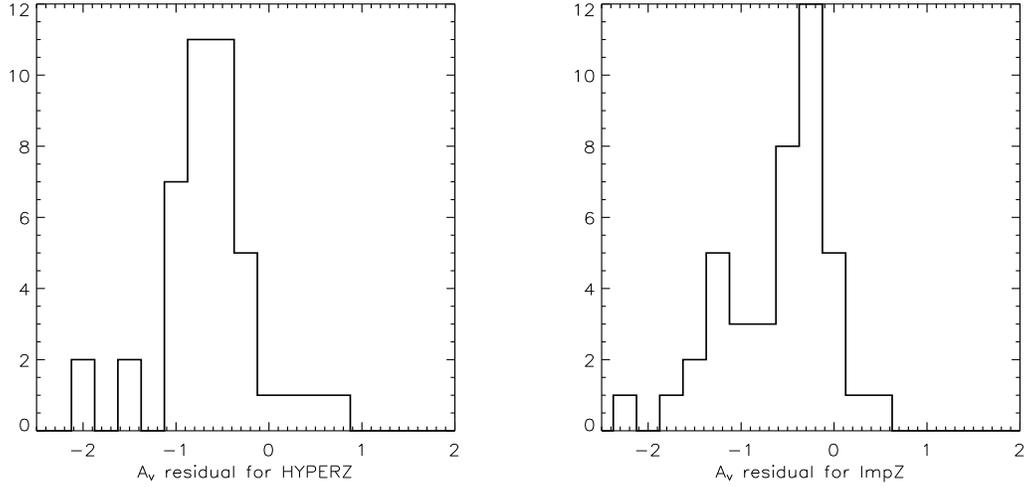}
\caption{ {\bf A$_{\rm V}$ residuals for HYPERZ (left panel) and I{\scriptsize MP}Z (right panel) with $\gamma$=0.25}.
The residual is now (0.25*Balmer[A$_{\rm V}$]-Phot[A$_{\rm V}$]).
}\label{fig:avresid08}
\end{center}
\end{figure*}

As before, the distribution of A$_{\rm V}$ residuals is plotted
(Figure \ref{fig:avresid08}).  It can be seen that the
distribution is more peaked, though offset from zero.

\begin{figure*}
\begin{center}
\includegraphics[width=15cm,height=7cm]{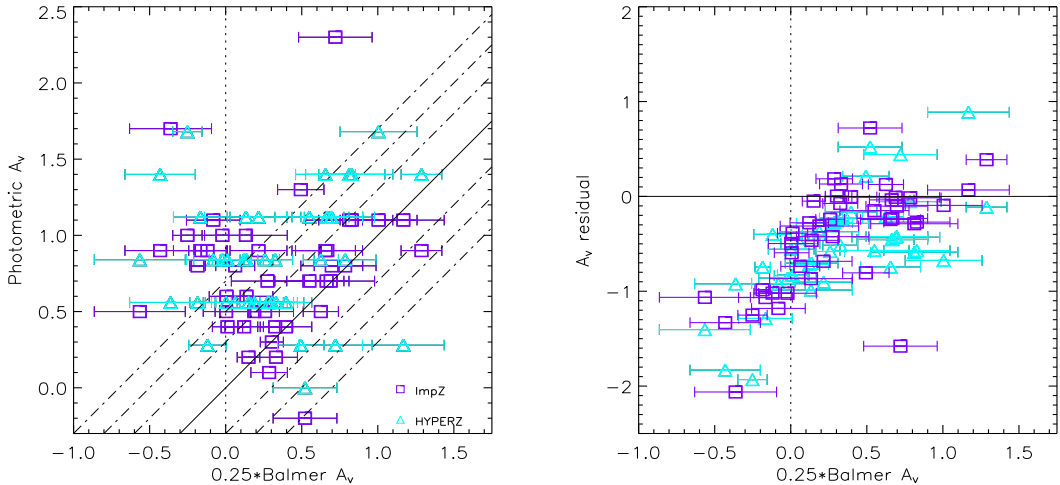}
\caption{ {\bf Left:} A$_{\rm V}$ results for HYPERZ (cyan
triangles) and I{\scriptsize MP}Z (purple squares).  Plot is the Balmer[A$_{\rm
V}$] (multiplied by $\gamma=0.25$) versus the
photometrically--derived A$_{\rm V}$.  Dot--dashed lines denote
residuals of 0.3, 0.5 and 0.7 in A$_{\rm V}$.   Errors are not
defined for the Phot-A$_{\rm V}$ values.  {\bf Right:}
A$_{\rm V}$ residuals for HYPERZ (cyan triangles) and I{\scriptsize MP}Z (purple
squares).  Plot is Balmer-A$_{\rm V}$ (multiplied by
$\gamma=0.25$) versus the residual (0.25*Balmer[A$_{\rm
V}$]-Phot[A$_{\rm V}$]).
}\label{fig:av_comp08}
\end{center}
\end{figure*}

Figure \ref{fig:av_comp08} plots the Phot-A$_{\rm V}$ values and
residuals as a function of Balmer-A$_{\rm V}$ (multiplied by
$\gamma$=0.25).  It can be seen that for lower Balmer-A$_{\rm V}$,
the Balmer-A$_{\rm V}$ tends to underestimate the extinction in
comparison to the Phot-$A_{\rm V}$ value.  If the negative Balmer-A$_{\rm V}$ 
are excluded, then the remaining sources with positive Balmer-A$_{\rm V}$
do tend to follow the line denoting agreement, albeit with large scatter.

\begin{figure*}
\begin{center}
\includegraphics[width=10cm,height=9cm]{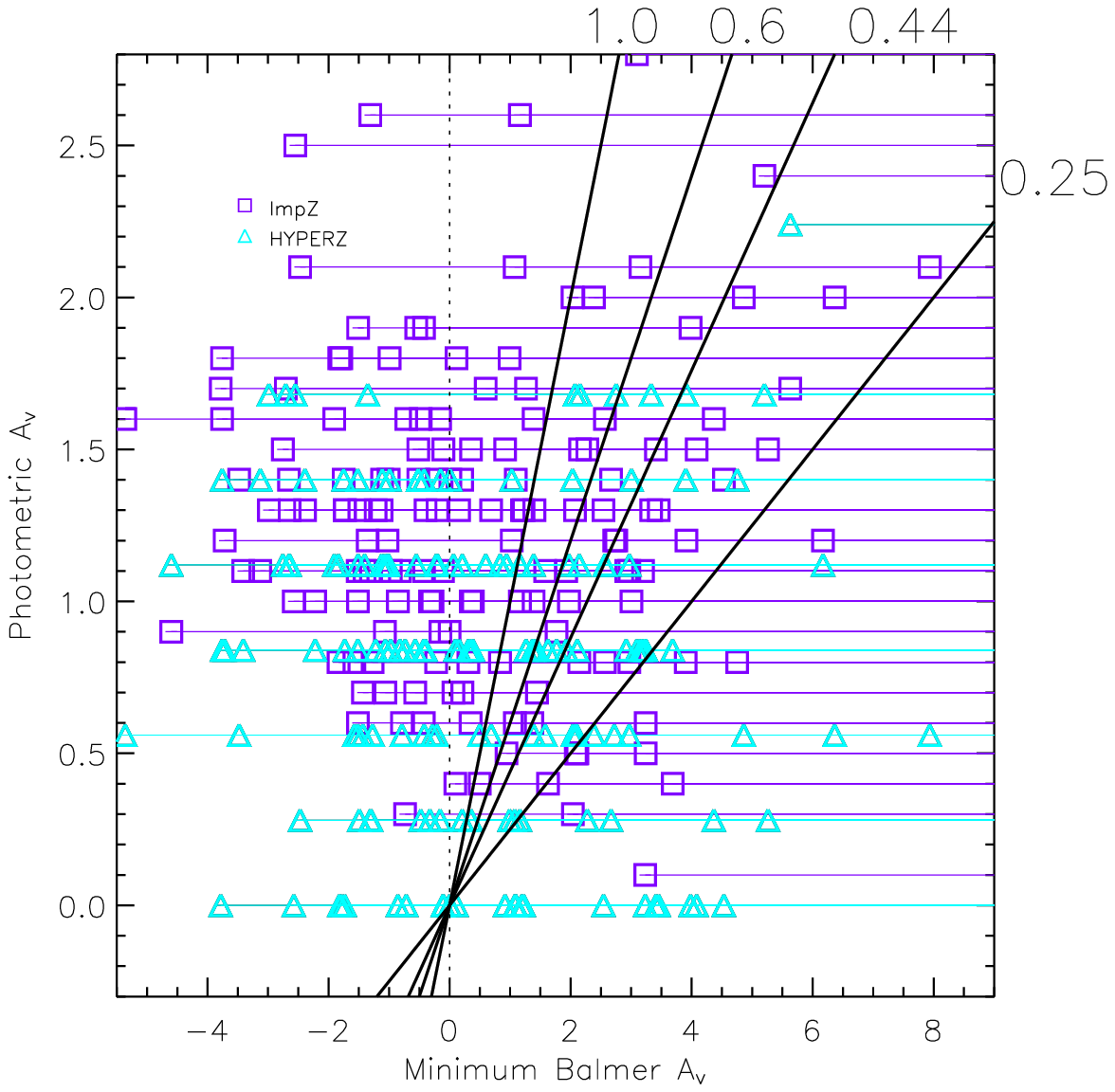}
\caption{ {\bf `Limit' cases:} A$_{\rm V}$ results for HYPERZ
(cyan triangles) and I{\scriptsize MP}Z (purple squares).  Plot is the
Balmer-derived A$_{\rm V}$ `limits' versus the
photometrically--derived A$_{\rm V}$.  Solid black lines denote
$\gamma$ values of 0.25, 0.44, 0.6 and 1.0 for the conversion
factor relating Balmer-A$_{\rm V}$ and Phot-A$_{\rm V}$.  Sources
need to lie on, or to the left of a line to imply consistency with
that chosen ratio.
}\label{fig:av_limits}
\end{center}
\end{figure*}

In Figure \ref{fig:av_limits} the sources with only a lower limit
on their Balmer-derived extinction (that is, a minimum 3$\sigma$
H$\alpha$ detection but only a limit on the H$\beta$ line
detection) are plotted in comparison to the extinction as derived
from the photometric redshift codes.  Here, no ratio $\gamma$ is
applied to the Balmer A$_{\rm V}$.  Instead straight lines
indicating different ratios are overplotted.  Since these are
lower limits, sources need to lie on, or to the left of a line to 
imply consistency with that chosen ratio.  It can be seen that
these lower limit sources are more consistent with lower values of
$\gamma$.  Of the 153 such sources, 146 (95\%) are consistent with the 
$\gamma$=0.25 line when considering I{\scriptsize MP}Z solutions (purple squares),
 but only 130 (85\%) are consistent when considering the HYPERZ solutions 
 (cyan triangles).
%
\section{Discussions and conclusions}
\label{sec:avdisc_conc}
A number of redshift codes routinely
output an A$_{\rm V}$ value in addition to the best-fitting
redshift, but little has been done to investigate the reliability
and/or accuracy of such extinction measures.  The main reason for
this lies in the aim of such codes - they have been developed and
optimised in order to derive {\it redshifts}.  However, as the
field of photometric redshift derivation matures, it is useful to
consider some of the other parameters that redshift solutions
output.

In this paper we have asked the question `Can a photometric
redshift code reliably determine dust extinction?'.  The short answer would be: 
`Not to a great accuracy'.\\

Using a sample with extinctions derived from Balmer flux-ratios the
A$_{\rm V}$ values produced by two photometric redshift codes,
I{\scriptsize MP}Z and HYPERZ, have been compared to the Balmer-A$_{\rm V}$
values.

First, it was demonstrated that the inclusion of A$_{\rm V}$ was crucial in order to 
obtain photometric redshifts of high accuracy and reliability, such that 95\% of the I{\scriptsize MP}Z 
results agreed with the spectroscopic redshifts to better than 0.1 in log[1+z]).  Without 
the inclusion of A$_{\rm V}$ freedom, there was a systematic and incorrect offset to higher 
photometric redshifts with many more incorrect redshift solutions.  The existence of some negative A$_{\rm V}$ solutions may be indicative of the need
for a bluer template in the template set, or for the inclusion of some additional free parameter in the fits.  
As the most important feature for 
template-fitted photometric redshifts is the location and identification of `breaks' in the SED, the inclusion of
A$_{\rm V}$ freedom can be seen, at first-order, as a modifier of the template SED's slope 
without having a strong effect on the
breaks themselves.   Hence a similar improvement may be achievable via a `tilting' parameter or similar which
would act to alter the slope of the template SEDs.  Since addition of dust extinction has a physical basis, however, 
this is a preferable parameter, as long as we can demonstrate that there is some correlation between the 
best-fitting phot-A$_{\rm V}$ and the actual (or in this case, that measured via the Balmer ratio) dust extinction of the source.  Thus, once the ability to derive good redshifts for the sample had been demonstrated a comparison between the phot-A$_{\rm V}$ and Balmer-A$_{\rm V}$ was carried out:

The correlation between the Phot-A$_{\rm V}$ and the
Balmer-A$_{\rm V}$ was similar for both codes, but was both noisy and 
not particularly strong.  Based on direct comparison between the two 
codes, and investigations into the $\chi^2$ solution space, a good part of this 
noise is derived from the inherent lack of precision the Phot-A$_{\rm V}$ solution 
has (perhaps 0.3 in A$_{\rm V}$ say), no doubt since it is based on
 only five photometric measurements.  Additional noise arises from the 
 precision of the Balmer-A$_{\rm V}$, typically accurate to perhaps 30\%, 
 which is due to the resolution of the spectrographic data.  Given these errors, 
 the correlation seen was in fact quite good.

The correlation was improved somewhat when the empirical 
value of $\gamma=0.44$, the
ratio between gas- and star- derived extinction, as determined by
\cite{Calzetti2001NewAR..45..601C}, was allowed to vary.  From
least-squares-fitting the minimum in the reduced $\chi^2$
distribution was found for $\gamma\sim0.25\pm0.2$.\\

The Calzetti ratio of 0.44 means that there is around a factor two
difference in reddening such that the ionised gas (as measured by
the Balmer decrement) is twice as reddened as the stellar continua
(as measured by the photometry) (e.g
\citealt{Fanelli1988ApJ...334..665F};
\citealt{Calzetti1994ApJ...429..582C}).  This implies that the
covering factor of the dust is larger for the gas than for the
stars, which can be explained by the fact that the ionising stars
are short-lived and so for their lifetime remain relatively close
to their (dusty) birthplace, whilst the majority of stars,
contributing to the galaxy's overall optical luminosity are
longer-lived and can migrate away from their dusty origins.

For the sample of galaxies in this paper, this factor two
difference in covering factor implied by the Calzetti ratio is
found to be plausible, given the errors of the method.  The sample
has a some preference for an increased covering factor which would
imply they are undergoing more rapid, `bursty' star formation than
the galaxies Calzetti used in her derivation. Perhaps more
importantly, the results demonstrate the pitfalls of assuming that
star- and gas-based extinction measures will give the same dust
extinction given some conversion factor.  Thus, correlation to
Balmer-derived values are modulo the uncertainty in
comparing star- and gas-based extinction measures.\\

However the results presented here show that, given certain
considerations, there is potential in the application of
photometric codes to reliably derive an extinction measure, though the precision
is currently low.  It is expected that the ability of photometric redshift codes to
determine extinction will improve with the availability of more
photometric bands (here, there are five wide band filters between
3,000-9,000{\AA}).  A sample with a combination of wide and
narrow-band filters, with good wavelength coverage and range (in
particular, extension to near-IR) will break many of the
degeneracies and allow the codes to accurately differentiate
between different possible fits.

The results also show that it is important to note that this will
be a measure of the star-based extinction, and will not
necessarily be well correlated with the extinction to the ionised
regions of a galaxy.
\section{Acknowledgements}
\label{sec:ack}
We would like to thank Michael Rowan-Robinson for discussions on
the nature of dust extinction and SED templates.  The referee provided astute 
suggestions and comments on this work and we extend our thanks.  We also thank
those responsible for the CNOC2 survey whose data we have used
here.  
\bibliography{/Users/tsb1/Documents/Work/References/BibdeskBibliog}
\bibliographystyle{mn2e}
\end{document}